# Visualizing high-temperature spin dynamics in $La_{1-x}Ca_xMnO_3$ from a mapping of EPR linewidth and $g$ factor

Y. Liu[*] and S. L. Wan

*Department of Modern Physics, University of Science and Technology of China,*

*Hefei 230026, P. R. China*

X. G. Li

*Hefei National Laboratory for Physical Sciences at Microscale, Department of Physics,*

*University of Science and Technology of China, Hefei 230026, P. R. China*

Abstract

We report the measurements of electron paramagnetic resonance (EPR) on powder samples of $La_{1-x}Ca_xMnO_3$ at the commensurate carrier concentrations of $x$ = N/8 (N = 1, 2, 3, 4, 5, 6, and 7) within temperature range $100K \leq T \leq 450K$. It is suggested that the mapping of EPR linewidth and $g$ factor is a useful way to reveal the high-temperature spin dynamics in $La_{1-x}Ca_xMnO_3$. An electron-hole asymmetry can be clearly observed in the mapping of $g$ factor. The linewidth $\Delta H$ as a function of $x$ drops at $x$ = 3/8 due to the mechanism of exchange narrowing. The activation energy $\Delta E$ obtained by fitting the intensity $I$ to an Arrhenius law displays a sharp peak at $x$ = 3/8, suggesting a strong ferromagnetic coupling. Our results convince that the spin-only relaxation mechanism should dominate the high-temperature paramagnetic regime in colossal magnetoresistance (CMR) manganites.

---

[*] Corresponding author. Email address: yliu@ustc.edu.cn

Due to a complex interplay among charge, spin, orbital, and lattice degrees of freedom, doped perovskite manganites $R_{1-x}A_x MnO_3$, where R is a trivalent rare earth ion and A is a divalent alkaline earth ion, show a rich phase diagram as a function of doping, temperature, and magnetic field [1-3]. The prototypical manganite system is $La_{1-x}Ca_x MnO_3$ (LCMO), which exhibits various ground states with the variation of carrier concentration. The phase diagram of LCMO was studied through the measurements of magnetization, resistivity, thermal conductivity, and thermopower by Cheong and co-workers in detail [4]. It is generally accepted that the competition between the double exchange (DE) interaction [5] and an enhanced electron-phonon coupling via the Jahn-Teller (JT) active $Mn^{3+}$ ion [6] plays a key role in determining the phase diagram of the manganites. A qualitative picture to explain the phase diagram of LCMO can be summarized: In samples with $0.2 < x < 0.5$, a transition from paramagnetic insulating (PI) to ferromagnetic metallic (FM) state occurs upon cooling due to the predominant DE mechanism. For the higher doping range with $0.5 < x < 0.875$ the ground state favors a charge ordering (CO) state. It was shown that the dominant mechanism responsible for the charge order is the JT coupling, with a lesser but significant contribution from the on-site Coulomb interaction [7]. In contrast to rich doping dependent ground states, high-temperature PI regime above the ordering temperatures seems to be simply dominated by the self-trapped small polarons [6,8]. X-ray and neutron scattering measurements have directly demonstrated the presence of short-range polaron correlations in the PI phase of optimally doped manganites [9-11]. Surprisingly, the ground state of $La_{5/8}Ca_{3/8}MnO_3$ is FM phase, but the insulating nature of the system above the Curie temperature $T_C$ is correlated with nano-scale charge/orbital ordering [12]. The results strongly suggest that the PI regime in LCMO phase diagram is more complicated than one's intuition thinking.

From the experimental point of view, much effort has been devoted to understand magnetic, electrical transport, and thermal properties in the paramagnetic state [8]. In these studies magnetic correlations are supposed to be completely ignored, since electron-phonon and on-site Coulomb interactions dominate in the high-temperature regime. However, one can not exclude the role of magnetic correlations when considering short-range polaron correlations. An analysis of the spin-spin correlations based on the Monte Carlo calculations shows that the ferromagnetic clusters form with size to be three-to-four lattice spacings above $T_C$ [13]. The short-range CO correlation at high temperature is possibly in the form of a FM "zigzag," a small segment of the CE-type CO state [12].

The main purpose of this study is to depict spin dynamics in the PI regime of LCMO phase diagram. The electron paramagnetic resonance (EPR) is a powerful probe of spin dynamics in the manganites. We report a systematic investigation of the temperature dependences of the EPR line resonance field ($g$ factor), linewidth $\Delta H$, and intensity $I$ for polycrystalline samples of LCMO at the commensurate carrier concentrations of $x = N/8$ (N = 1, 2, 3, 4, 5, 6, and 7). As is evident in the phase diagram, there are well defined features at the commensurate carrier concentrations of $x = N/8$ (N = 1, 3, 4, 5, and 7) in LCMO [4], e.g., optimal doping for FM state at $x = 3/8$, the highest CO temperature $T_{CO}$ at $x = 5/8$. Moreover, Curie temperature $T_C$ is similar to the temperature $T_{CO}$ for the two doping levels, forming an electron-hole symmetry phase line centered at $x = 4/8$.

High-quality polycrystalline samples LCMO with commensurate doping of $x = N/8$, (N = 1~7) were prepared by a standard solid-state reaction with an identical synthesis condition. Stoichiometric proportions of $La_2O_3$, $CaCO_3$, and $MnCO_3$ were mixed and heated at 1200 °C for two days with an intermediate grinding. After grinding, the mixture was pressed into pellets and sintered at 1300 °C for 24 h. Phase purity and crystal structure of the samples were characterized by x-ray diffraction (XRD). Figure 1 shows the Ca doping dependence of XRD patterns for LCMO ($x = N/8$, N = 1~7) at room temperature, which could be indexed in a *Pnma*-type orthorhombic structure. The EPR spectra were recorded using a Bruker ER200D spectrometer at 9.61 GHz (*X* band) upon warming within the temperature range $100K \leq T \leq 450K$. The measurements were carried out on loose-packed micron-sized crushed crystals.

Figure 2 shows the EPR spectra recorded as the derivative *dP/dH* at room temperature for the samples. Each ESR spectra consists of a symmetric Lorentzian line. Oseroff *et al*. [14] suggested that the EPR signal in the manganites is a consequence of magnetic clusters made of a collection of $Mn^{3+}$ and $Mn^{4+}$ ions. In the EPR measurements, the parameters of primary interest are the *g* factor and the linewidth $\Delta H$. Since the lines are very broad both in powder and single crystals, for accurate determination of the various line shape parameters we have fitted the signals to appropriate line shape functions. For powder samples small compared to the skin depth, one expects a symmetric absorption spectrum [15]

$$\frac{dP}{dH} \propto \frac{d}{dH}\left(\frac{\Delta H}{(H-H_r)^2+\Delta H^2}+\frac{\Delta H}{(H+H_r)^2+\Delta H^2}\right), \qquad (1)$$

where $H_r$ is the resonance field and $\Delta H$ is the linewidth. As shown in Fig. 2, the fits of the signals to Eq. (1) are excellent.

From the best fit value of the resonance field $H_r$, the *g* value is obtained from the resonance condition: $h\nu = g\mu_B H_r$. Figure 3(a) plots the temperature dependence of *g* value for the samples. It is found that the *g* values are nearly temperature independent except those close to the ordering temperatures. Interestingly, *g* values divide into two distinct parts for the hole doped and electron doped samples. By linearly interpolating the results, we can draw a false color mapping of *g* factor in the *T* vs *x*, as shown in Fig. 3(b). It is noteworthy that *g* factor displays an electron-hole asymmetry forming a phase line centered at *x* = 4/8. The feature of electron-hole asymmetry was also observed in $Pr_{1-x}Ca_xMnO_3$ (PCMO), where the *g* value for the electron doped sample (*x* = 0.64) is less than the free electron value $g_e$ ~2.0023, whereas for the hole doped one (*x* = 0.36) it is more than $g_e$ at room temperature [16]. In powder samples the individual grains are randomly orientated. The local field has little influence on the resonance field, leaving the $g$ value unaffected. Thus we believe the electron-hole asymmetry of $g$ factor is an intrinsic phenomenon.

In an octahedral anion crystal field $g$ factor for $3d^3$ ions can be expressed by $g = g_e - 8\lambda/\Delta$, where $\lambda$ is the spin–orbit coupling constant, $\Delta$ is the energy gap between the ground level and the excited level in question [17]. In the manganites the transition from the ground state to the excited state corresponds to promoting an electron from a $t_{2g}$ to an $e_g$ orbital, where the splitting of the energy levels is ~1.5 eV [18]. Thus, the values of $\lambda$ are estimated as 64 cm$^{-1}$ and 85 cm$^{-1}$ for the hole doped and electron doped LCMO, respectively. These values are strongly reduced compared with the data ~350 cm$^{-1}$ observed in individual $Mn^{3+}$ and $Mn^{4+}$ ions [19]. The large deviation may suggest that: (I) the spin–orbit coupling is strongly suppressed in the manganites; (II) the main contribution to EPR signal could not be attributed to $Mn^{4+}$ ions only. A theoretical approach suggests that the EPR line position in weakly anisotropic Heisenberg magnets can be described by three contributions: the first one, a static shift, comes from the magnetic susceptibility anisotropy, the second one, a dynamic shift, is due to the spin motion and comes from the self-energy, while the last one comes from the asymmetry of the line shape [20]. The electron-hole asymmetry of $g$ factor might result from the effect of dynamic shift. For $x < 4/8$ $e_g$ electrons can hop readily from one Mn site to another due to the predominant DE interaction. However, they tend to localize for $x \geq 4/8$, while superexchange interaction becomes active [4].

In contrast to $g$ factor, the linewidth $\Delta H$ for the samples show a wide variety of behaviors depending on both the temperature and doping level, as shown in Fig. 4(a). As a function of $T$, the EPR linewidth $\Delta H$ for all the samples except $x = 7/8$ decreases with decreasing temperature. The linewidth $\Delta H$ of the sample $x = 7/8$ saturates rapidly with increasing temperature. Upon cooling $\Delta H$ in all cases goes through a minimum at $T_{min}$. Below this temperature $\Delta H$ increases rapidly. It was reported that this behavior is not intrinsic but extremely sample dependent [21]. The feature can be well described in terms of a two-magnon scattering relaxation mechanism induced by the demagnetization fields of the pores between crystallites. Since the behavior observed between the ordering temperatures and $T_{min}$ is sample dependent and may confuse the analysis of the results in the paramagnetic regime, we have excluded this small temperature region from our analysis.

We show the mapping of the EPR linewidth in Fig. 4(b). Remarkably, on approaching the ordering temperatures from above, the narrowing of EPR linewidth gets strong for $x \leq 0.5$, where colossal magnetoresistance (CMR) effect is observed. It is found that the narrowing of EPR linewidth is strongly correlated with $T_C$. The linewidth $\Delta H$ is the narrowest for the optimally doped sample $x = 3/8$ at the same temperature. The drop of $\Delta H$ extends to the highest temperature, related to the highest Curie temperature $T_C$ at this doping level. Let's turn to the resistivity $\rho$ at 300 K for LCMO [4]. Note at 300 K the smooth behavior as $x$ grows from 0, only interrupted close to $x = 1$ when the G-type antiferromagnetic insulating (AFI) state is reached. It was found that, however, doping dependence of magnetic susceptibility $\chi$ at 300 K display a strikingly sharp peak at $x = 3/8$ composition [12]. According to Zener's DE mechanism [5], $e_g$ electrons hop between $Mn^{3+}$ and $Mn^{4+}$ ions while keeping their spin directions due to a strong Hund coupling energy because such hopping is most probable when the spins of $t_{2g}$ electrons of the $Mn^{3+}$ are aligned with those of the adjacent $Mn^{4+}$. Thus, the enhancement of FM correlation is related to the strongest exchange interactions at $x = 3/8$.

It is well known that in a paramagnet where the nearest neighbouring spins are coupled by an exchange interaction $J$, the EPR signal should be strongly exchange narrowed so that instead of the full dipolar width $\Delta H_0$ [22]. One should observed a linewidth

$$\Delta H = (\Delta H_0)^2 / \Delta H_{ex} \qquad (2)$$

where $\Delta H_{ex}$ is the exchange field. $\Delta H_{ex}$ is quite large and related to the Curie temperature

($S\mu_B \Delta H_{ex} \approx k_B T_C$). Thus the drop of linewidth $\Delta H$ at $x = 3/8$ can be roughly interpreted in terms of exchange narrowing when leaving the dipolar fields unchanged [23]. It has been found that the polaronic state near the optimal doping is intrinsically inhomogeneous, consisting of magnetic clusters 10 to 20 Å in diameter [9-12]. Generally, the polaronic state is a consequence of the strong electron-phonon coupling, enhanced by the JT activity of $Mn^{3+}$ ion in the manganites. On the other hand, the exchange correlation is strongly enhanced for $x$ near 3/8, which is relevant for the presence of short-range polaron correlations. Our observation naturally suggests that the exchange correlation provides the "glue" for the formation of FM coupled polarons. This result presents great implications on the origin of the EPR signal in the manganites. It seems reasonable to identity short-range correlated polarons with the spin entity suggested by Oseroff *et al*. [14].

It is still a controversial issue on the relaxation mechanism for the temperature dependence of linewidth $\Delta H$ in the manganites. The linewidth $\Delta H$ shows a linear $T$ dependence between $1.1 T_C < T < 2 T_C$ for hole doped manganites, which was interpreted in terms of a single-phonon spin-lattice relaxation mechanism [23,24]. By substituting $^{16}O$ for $^{18}O$, the characteristic differences observed in EPR intensity and linewidth for the two isotope samples were suggested to be caused by a bottlenecked spin relaxation takes place from the exchange-coupled constituent $Mn^{4+}$ ions via the $Mn^{3+}$ Jahn-Teller ions to the lattice [25]. Further, Shengelaya *et al*. [26] found that the temperature dependence of linewidth $\Delta H$ can be described by the adiabatic hopping of small polarons, *i.e.*, $\sigma \propto T^{-1} \exp(-E_a / k_B T)$, which is consistent with the existence of a bottleneck EPR regime in the manganates. We have fitted our data show in Fig. 4(a) by the following expression

$$\Delta H = \Delta H_0 + \frac{A}{T} \exp(-E_a / k_B T), \qquad (3)$$

where $E_a$ is the activation energy, i.e., the potential barrier that the polaron must surmount in order to hop into the next site. The $E_a$ values obtained by fitting Eq. (3) are plotted in Fig. 5(a). The data obtained in the previous work at $x$ = 0.18, 0.2, and 0.22 are also presented in Fig. 5(a) [26,27]. It is found that $E_a$ values display a peak at $x$ = 3/8. However, $E_a$ values deduced from the conductivity measurements show a striking divergence from those measured by EPR technique. As shown in Fig. 5(a), they decrease smoothly with increasing $x$ [28]. No anomalous behaviors are observed at $x$ = 3/8. As can be seen in Fig. 5(a), the coincidence of $E_a$ values only occurs around $x$ = 1/3 and 5/8. On the other hand, Atsarkin *et al*. [29] have measured the longitudinal spin-relaxation time $T_1$ in the paramagnetic state of three LCMO samples ($x$ = 0.2, 0.25, and 0.33). The reported $T_1$ behavior contradicts that predicted by the polaron model.

According to Ref. 25, the EPR signal observed in LCMO is due primarily to $Mn^{4+}$ ions. It was found that, however, both $Mn^{4+}$ and $Mn^{3+}$ ions take part in producing the EPR signal [23]. A further study suggest that EPR susceptibility $\chi_{esr}$, deduced from the relation $I \propto \chi_{esr}$ ($I$ is EPR intensity), could be identified with magnetic susceptibility $\chi_{dc}$ [30]. The EPR experiment probes the dynamics of the magnetic system. The coincidence of $\chi_{esr}(T)$ with $\chi_{dc}(T)$ in the whole PI range studied clearly indicates that all the Mn ions contribute to the observed EPR spectra. Thus, the EPR linewidth should be related to the relaxation mechanism of the coupled magnetic system. In this study, the EPR intensity $I$ was determined by numerical double integration of the measured spectra. Instead of a simple Curie-Weiss law, the intensity $I$ during PI regime follows an Arrhenius law [13,14,30],

$$I = I_0 \exp(\Delta E / k_B T), \quad (4)$$

where $\Delta E$ is the activation energy of spin clusters. Since the EPR signal is associated with some form of the magnetic cluster [14], the $\Delta E$ would be required in order to dissociate these spin entities made of collection of individual spins [13]. Figure 5(b) shows $\Delta E$ obtained by a linear fit of ln $I$ vs 1000/$T$ plots. Interestingly, the activation energy $\Delta E$ resembles the result obtained by fitting linewidth Δ*H* with Eq. 3, and peaks at *x* = 3/8. This feature was also observed by Oseroff *et al*. [14]. The peaking behavior of Δ*E* should be related to the strong FM coupling at *x* = 3/8 [12].

In any case, the nature of the EPR broadening is determined by the mechanism of the electron spin relaxation. According to Refs. 30-32, the EPR linewidth in a wide variety of the perovskite manganites is determined by spin-spin (exchange) interactions between the $Mn^{3+}$ and $Mn^{4+}$ ions and so is unrelated to any spin-lattice processes. In all cases studied, the linewidth Δ*H* away from magnetic and structural transitions can be fitted to a simple expression [30-32]

$$\Delta H = [C/T\chi_{dc}(T)]\Delta H(\infty), \qquad (5)$$

where Δ*H*(∞) is system dependent constant, and may be identified with the high-temperature limit of the linewidth. Using the relation $I \propto \chi_{esr} \sim \chi_{dc}$, one can easily deduce that the production of Δ*H*×*I* is in proportion to the inverse temperature. Figure 6 shows the plots of Δ*H*×*I* vs 1000/*T*. Departing from the critical regime, a linear behavior is clearly observed. The result is in good agreement with spin-only relaxation mechanism.

In summary, we demonstrate that the mapping of EPR parameters offers a powerful tool to investigate high-temperature spin dynamics in the phase diagram of CMR manganites. An electron-hole asymmetry of *g* factor is linked to the dynamic effect of $e_g$ electron motion. The narrowing behavior of linewidth Δ*H* and peaking behavior of activation energy Δ*E* as a function of *x* reveal the strong FM coupling at *x* = 3/8. The analysis of linewidth Δ*H* and intensity *I* supports that the EPR signal originates from the magnetic clusters, dominated by spin-spin exchange interaction.

This work was supported by the National Nature Science Foundation of China and the National Basic Research Program of China.

**Figure captions**

Fig. 1 The Ca doping dependence of XRD patterns for polycrystalline samples of $La_{1-x}Ca_xMnO_3$ ($x$ = N/8, N = 1, 2, 3, 4, 5, 6, and 7) at room temperature.

Fig. 2 EPR spectra of powder samples of $La_{1-x}Ca_xMnO_3$ ($x$ = N/8, N = 1, 2, 3, 4, 5, 6, and 7) at 300 K. The solid lines show the fits of the experimental data to Eq. (1).

Fig. 3 (a) Temperature dependence of the resonance field ($H_r$) of EPR signals presenting by $g$ factor for $La_{1-x}Ca_xMnO_3$ samples ($x$ = N/8, N = 1, 2, 3, 4, 5, 6, and 7). (b) A false color mapping of $g$ factor in the $T$ vs $x$ plane for $La_{1-x}Ca_xMnO_3$. The solid magenta circles show $T_{min}$'s for the measured compositions, and red line is guide to the eye. Dashed line indicates the boundary of two different $g$-value regimes at $x$ = 4/8.

Fig. 4 (a) Temperature dependence of the linewidth $\Delta H$ for $La_{1-x}Ca_xMnO_3$ samples ($x$ = N/8, N = 1, 2, 3, 4, 5, 6, and 7). The solid lines correspond to the best fits of Eq. (3). (b) A false color mapping of linewidth $\Delta H$ in the $T$ vs $x$ plane for $La_{1-x}Ca_xMnO_3$. The solid magenta circles show $T_{min}$'s for the measured compositions, and red line is guide to the eye.

Fig. 5 (a) Doping behavior of the activation energy $E_a$ extracted from temperature dependent EPR linwidth and resistivity for $La_{1-x}Ca_xMnO_3$ system using the adiabatic small polaron model [Eq. (3)]. Solid circle corresponds to the data at $x$ = 0.2 obtained by Shengelaya *et al*. [26]. Solid triangles correspond to the data at $x$ = 0.18, 0.2, and 0.22 reproduced from Ref. 27. The data obtained by electronic measurements are adapted from Ref. 28. (b) Activation energy $\Delta E$ obtained by the fits of double integrated intensity of EPR signals with the Arrhenius law [Eq. (4)].

Fig. 6 The plots of $\Delta H \times I$ vs 1000/$T$ showing a linear behavior in the whole doping range.

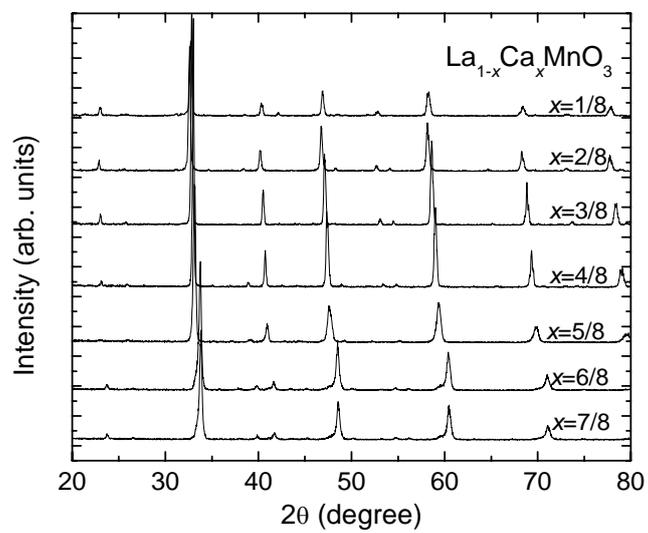

Fig. 1 by Y. Liu *et al*.

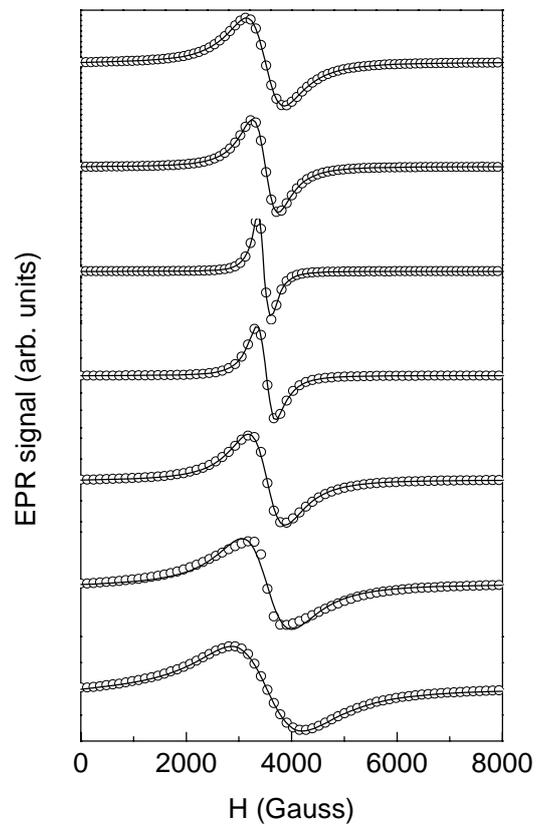

Fig. 2 by Y. Liu *et al*.

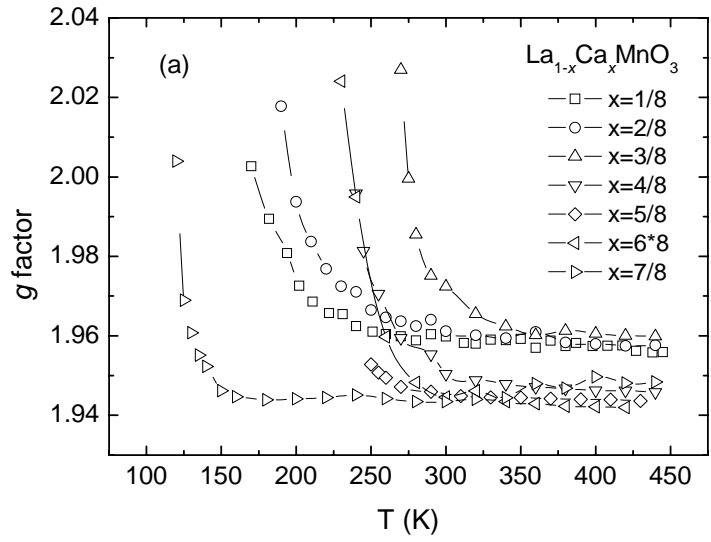

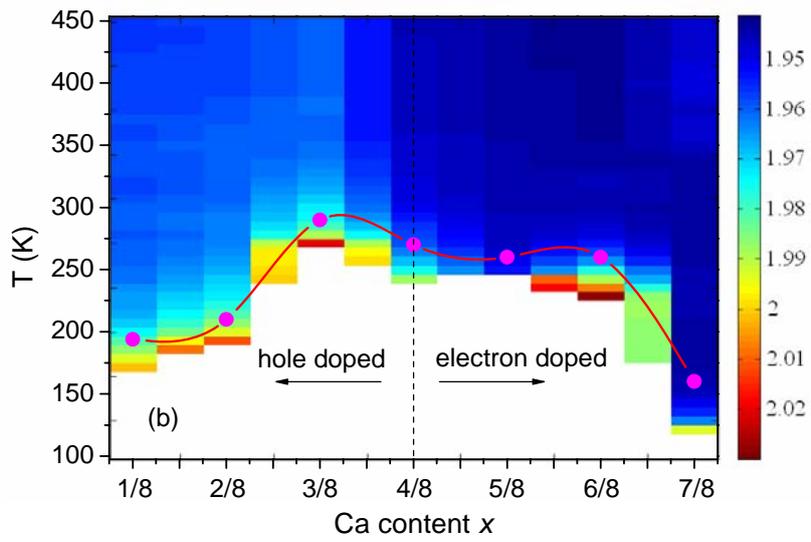

Fig. 3 by Y. Liu *et al*.

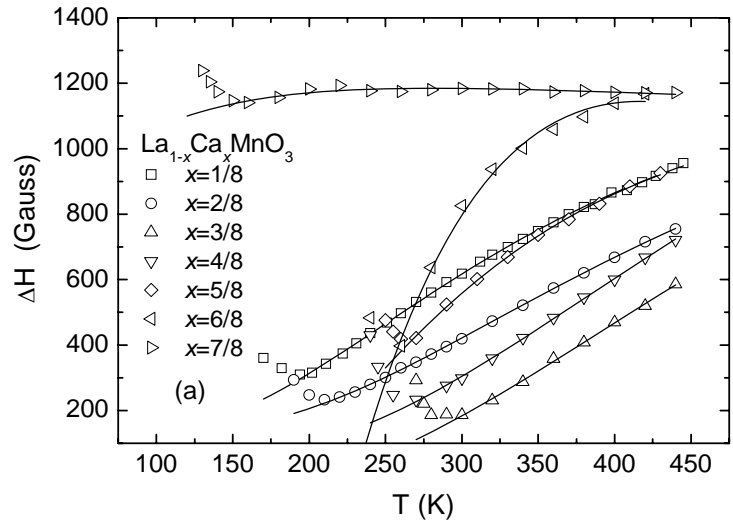

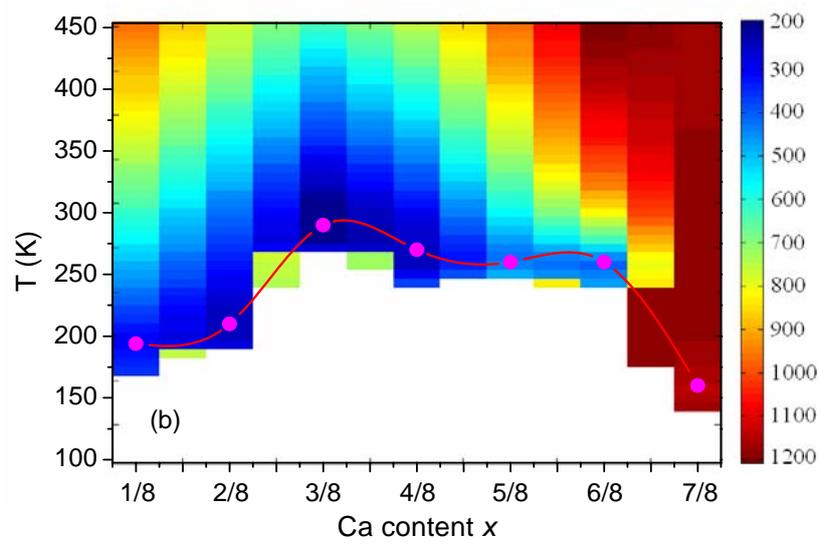

Fig. 4 by Y. Liu *et al*.

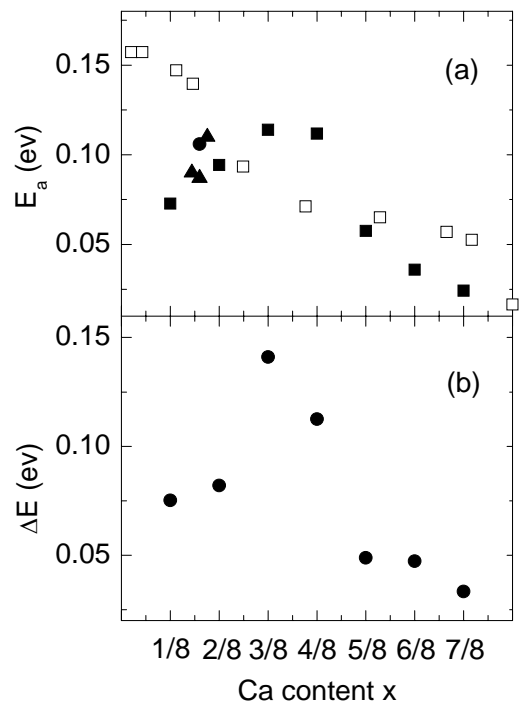

Fig. 5 by Y. Liu *et al*.

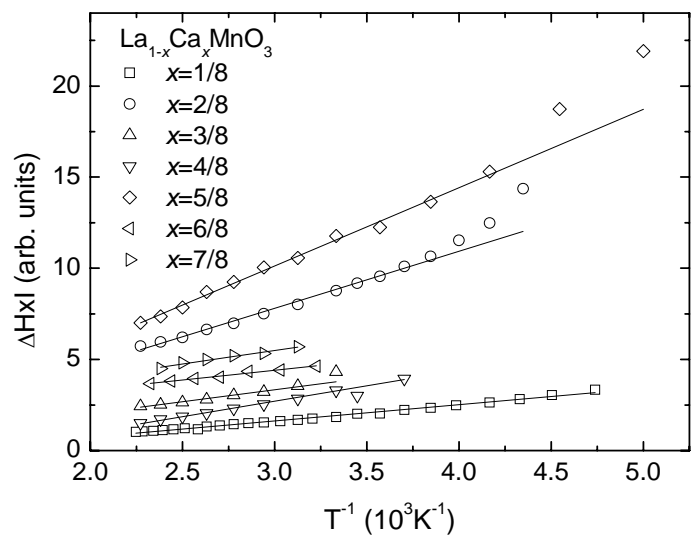

Fig. 6 by Y. Liu *et al*.